\documentclass{osa-article}
\pdfoutput=1

\journal{osajournal}

\articletype{Research Letter}

\usepackage{lineno}
\usepackage[utf8]{inputenc}
\usepackage[T1]{fontenc}
\usepackage{textgreek}
\usepackage{hyperref}
\usepackage{newtxtext,newtxmath}

\begin{document}

\title{Spatiotemporal refraction of light in an epsilon-near-zero ITO layer}

\author{Justus Bohn\authormark{1,*},Ting Shan Luk\authormark{2,3}  and Euan Hendry\authormark{1}}

\address{\authormark{1} School of Physics, University of Exeter, Exeter, UK\\
\authormark{2}Sandia National Laboratories, Albuquerque, NM, USA\\
\authormark{3}Center for Integrated Nanotechnologies, Sandia National Laboratories, Albuquerque, NM, USA}

\email{\authormark{*}jb933@exeter.ac.uk} 



\begin{abstract}
When light travels through a medium in which the refractive index is rapidly changing with time, the light will undergo a shift in its frequency. Large frequency shifting effects have recently been reported for transparent conductive oxides. These observations have been interpreted as emerging from temporal changes to the propagation phase in a bulk medium resulting from temporal variations in the refractive index, an effect referred to as temporal refraction. Here, we show that the frequency shift in an epsilon-near-zero (ENZ) layer made of indium tin oxide (ITO) originates not only from this bulk response, but includes a significant effect resulting from temporal changes to the spatial boundary conditions. This boundary effect can lead to a dominant opposing shift to the bulk effect for certain angles. Hence, this process gives rise to a frequency shift that can be tailored through angle, decoupling the amplitude and phase modulation.
\end{abstract}

\section*{Introduction}

All-optical signal processing requires the control of various parameters of light waves such as the amplitude, phase and frequency. Recent ambitions paint a promising picture for all-optical switching, showing sub-picosecond and large amplitude modulation for a variety of platforms, based on amorphous silicon \cite{Shcherbakov2015UltrafastAllOpticalSwitching}, gallium phosphide \cite{Grinblat2019UltrafastSub30fs}, plasmonic waveguides \cite{Ono2020UltrafastEnergyefficientAlloptical}, and epsilon near zero (ENZ) layers\cite{Kinsey2015,Caspani2016a,Yang2017}, including the here used indium tin oxide (ITO) \cite{Alam2016,Bohn2021AllopticalSwitchingEpsilonnearzero}. Even cavity based optical transistor have been demonstrated \cite{Chen2013AllOpticalSwitchTransistor,Zasedatelev2019}. However, while the amplitude of light can be straightforwardly controlled using these different modulator materials, controlling the frequency of light is more challenging. To modulate the frequency of a light signal, one can look to change the phase of a light signal on ultrafast time scales, effectively creating a temporal refractive index boundary \cite{Akhmanov1969NonstationaryPhenomenaSpacetime,AuYeung1983PhaseconjugateReflectionTemporal,Mendonca2000TheoryPhotonAcceleration}. However, in many circumstances changes to frequency are necessarily and directly coupled to changes in amplitude, making independent modulation difficult. 
 
Recently, thin films of epsilon-near-zero (ENZ) materials have offered a promising route to frequency modulation \cite{Bruno2019,Zhou2020,Khurgin2020b,Liu2021}. In such materials, large (> unity) changes to refractive index can be induced on sub-100\,fs timescales due to ultrafast heating of the electron gas, an effect which can result in frequency shifts by up to a few percent of the carrier frequency \cite{Alam2016,Caspani2016a}. However, while observed frequency shifts are large when temporal refraction occurs in these systems, they are also directly linked to refractive index changes which lead to simultaneous amplitude modulation, preventing independent optimization.

Here, we investigate spatiotemporal refraction for tailored frequency shifting in thin indium tin oxide layers. We show that the frequency shift arises not only from a bulk response, but includes a significant contribution from temporal changes to the spatial boundary conditions. The frequency shift arising from boundary effects can oppose that from the bulk effect, and lead to a strong contribution for layer thicknesses up to \textasciitilde 1\,\textmu m. We further show that for high incident angles, it gives rise to a dominant opposing shift, i.e. shifting to higher rather than the usual lower frequencies, while maintaining an increase in differential transmission. This competition between surface and bulk responses could have applications where tuning of the amplitude and direction of frequency shift is useful, or decoupling of amplitude and phase modulation is required.

\section*{Results}

\subsection*{ITO layer model}

For temporal refraction, a bulk medium of homogeneous refractive index undergoes a temporal change in refractive index throughout the bulk \cite{Donaldson2015WhatTemporalAnalog}. The temporal shape of the phase change is determined by the nonlinear refractive index change $\Delta n = n_2\,I$, which depends on the nonlinear material characteristics ($n_2$) and the pump pulse intensity ($I$). Time translation symmetry breaking leads to a change in frequency following \cite{Zhou2020} $n_1 f_1=n_2 f_2$  and subsequently $\Delta f=-\Delta n f_1/(n_1+\Delta n)$, which predicts that the frequency shift $\Delta f/f$ should depend only on the refractive index change $\Delta n/(n_1+\Delta n)$.

\begin{figure}[htb]
	\centering
	\includegraphics[width=0.6\columnwidth]{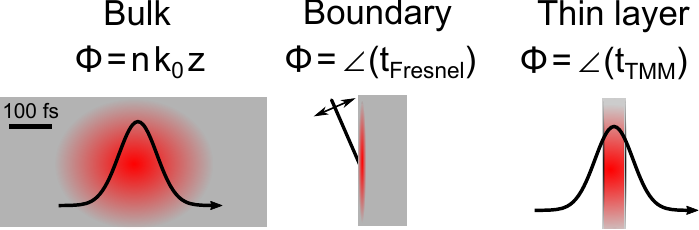}
	\caption{\label{fig:schematic} \textbf{Temporal changes in an optical medium.}  The schematic of the bulk propagation case highlights that the nonlinear refractive index facilitates a temporally changing phase for intense laser pulses. The Fresnel coefficients can be used to extract phase changes at a spatial boundary. Temporal changes to the interface induced phase-jumps also shift the frequency. Finally, the thin layer case studied in this paper combines both of these phenomena in the phase of the coefficient extracted via the transfer matrix method (TMM).}
\end{figure}

However, nonlinear media are becoming thinner, and films are normally considerably thinner than laser pulse lengths. This is particularly true for films of ITO, which are highly absorptive close to the ENZ frequency for films that are significantly thicker than a few hundred nanometers thickness. For such thin films, it is more useful to consider frequency shifts as arising from time dependent changes to the transmitted phase of a laser pulse. If we consider the change in phase arising from the propagation through the bulk of the film, $\Delta\Phi(t) = \Delta n(t)\,k_0\,L$ (see \autoref{fig:schematic}), we see that the thickness of the film is expected to limit the change in phase, and therefore the maximum observable frequency shift.

However, in thin films interfaces play an important role in determining transmission phase. The phase change upon crossing an interface is independent of the thickness. Effects arising from a temporal change of the Fresnel coefficients have been suggested to contribute to frequency shifting\cite{Bruno2020}. We start investigating these potential interface effects by considering the boundary conditions of the normal electric field component: $E_{\rm ITO}=E_{\rm air}/\varepsilon_{\rm ITO}$. When the complex ITO permittivity $\varepsilon_{\rm ITO}$ varies with time, the complex field in the ITO will also change with time, resulting in time dependent spatial refraction at the interface, referred to here as spatio-temporal refraction. The temporal change to this boundary induced phase can be associated with a frequency shift  $\Delta \omega = - {\rm d}\Phi/{\rm d}t $, expected to provide a thickness independent frequency shift besides the temporal refraction induced by the bulk.

\begin{figure}[htb!]
	\centering
	\includegraphics[width=0.7\columnwidth]{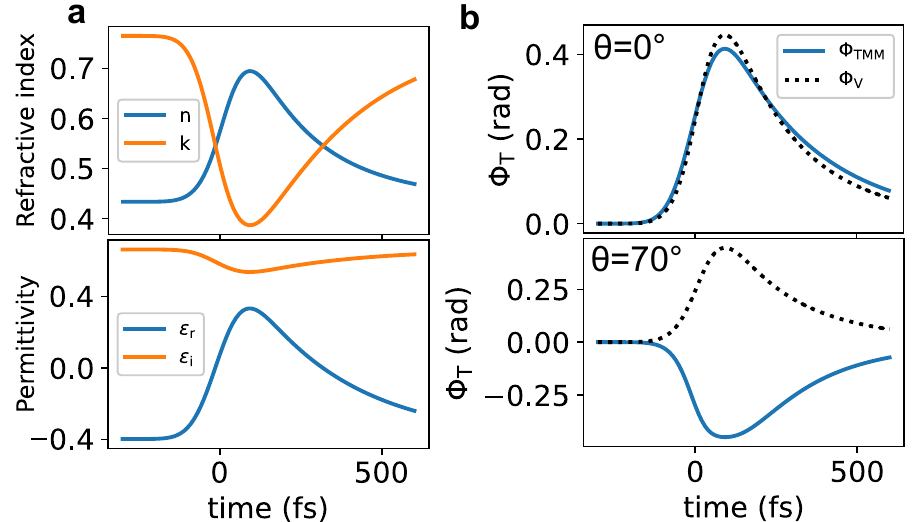}
	\caption{\label{fig:model} \textbf{Spatiotemporal refraction simulation.} We study the impact of a typical $\omega_p$ reduction by 10\,\% for the 407\,nm ITO sample parameters at 200\,THz. \textbf{a}, The refractive index $n$ initially increases until the electron gas reaches the minimum $\omega_p$, while the imaginary part $k$ decreases. They cross roughly at the maximum gradient, corresponding to the ENZ point as seen below for the permittivitty. The thermal decay is modelled with a time constant of 150\,fs. \textbf{b}, For the case of normal incidence the transmission phase simulated with the TMM ($\Phi_{\rm TMM}$) agrees very well with the bulk phase ($\Phi_{\rm V}$). However, for an incoming angle of $70^\circ$ a total thin layer phase shift of opposite sign is expected.}
\end{figure}

We simulate temporal changes to the ITO thin layer system via the transfer matrix method (TMM). Here, we treat the time dependent changes to permittivity as being homogeneous throughout the ITO layer. In supplementary section S3, we will revisit this assumption and show that spatial in-homogeneity due to the exponential decay of the pump throughout the sample leads to only minor quantitative effects. We consider an incident beam from air, followed by an ITO layer (optical properties in supplementary section S1), followed by the substrate (cover slip). Temporal changes to the ITO permittivity are defined by the convolution of our gaussian pump pulse, with a pulse length of 107\,fs, and an exponential decay of 150\,fs, determined from optical pump-probe experiments similar to \cite{Bohn2021AllopticalSwitchingEpsilonnearzero}. For a 407\,nm sample with the epsilon-near-zero frequency at $f_{\rm ENZ}=211.5\,$THz and a probe frequency $f_{\rm pr}=200\,$THz undergoing an estimated 10\,\% red-shift of the plasma frequency ($\omega_{\rm p}$), the time dependent optical properties are plotted in \autoref{fig:model}a. For a red shifting $\omega_{\rm p}$ the real part of the refractive index $n$ increases, while the imaginary part initially decreases on pumping. For the plotted example the real part of the permittivity crosses zero upon pumping, corresponding to the ENZ point.

 Using TMM, we extract the transmission coefficient of our thin layer based on the time dependent refractive index and plot the phase in \autoref{fig:model}b. The resulting time dependent phase for the normal case is  similar to that predicted from $\Delta\Phi_{\rm V}$, the change in phase expected due to propagation through the bulk - we discuss the differences between these two results in supplementary section S2.
 
However, increasing the angle of incidence leads to a very different phase response. As the angle of incidence is increased, the temporal gradient of the transmission phase can even change sign as seen for the example of $70^\circ$ in \autoref{fig:model}b. This behaviour arises due to the front interface, which shows a decrease in phase for the permittivity transitioning from negative to positive real values (see supplementary section S2.1). This effect is fundamentally different from the bulk response, demonstrating the importance of the interfaces in determining changes in phase and frequency. Interestingly, such a contrasting behaviour of the phase would result in an obvious experimental signature: a blue shift of the frequency as opposed to the red shift expected from a bulk. A more in-depth guide for understanding the model, the individual contributions and the different limits can be found in supplementary section S2.

\subsection*{Experimental measurement}
\begin{figure*}[htb!]
	\centering
	\includegraphics[width=0.95\linewidth]{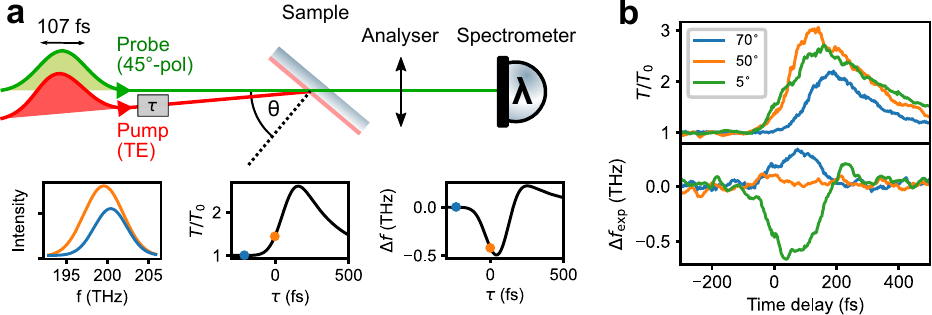}
	\caption{\label{fig:setup} \textbf{Experimental setup and measurement. a},  The setup consists of a TE polarized pump and a 45$^\circ$ probe that enables quick spectral measurement of either polarization by choice of analyser. The samples of interest are ITO films of 407\,nm or 115\,nm thickness on top of a coverslip. The lower plots show schematically how the measured spectra (left) at different time delays (blue, orange) can be used to extract changes in both transmission (middle) and central frequency (left). \textbf{b}, Measurements taken for three different angles. They correspond to an incoming pump intensity of $I_0=400$\,GW\,cm$^{-2}$,$f_{\rm pm}=250$\,THz, $f_{\rm pr}=200$\,THz and a TM analyser. The transmission increases up to $200$ to $300$\,\% for all angles, while the frequency shift is either negative ($5^\circ$), negligible ($50^\circ$) or positive ($70^\circ$).}
\end{figure*}  
To experimentally investigate this signature, we carry out a pump-probe measurement as depicted in \autoref{fig:setup}a. We use a pump intensity of 400\,GW\,cm$^{-2}$, resulting in frequency red shifts of $\sim1\,$THz. We note that our observed frequency shifts reported below are lower than those reported in \cite{Zhou2020}. This is due to a higher Drude scattering rate and a reduced thickness for our ITO samples. 

A TE pump polarization is chosen as it provides only a small angle and frequency dependence in absorption. We use a probe polarized at 45$^\circ$ to check frequency shifting behaviour of either polarization by rotating the analyser in front of the spectrometer. Spectra are recorded for different pump delay times $\tau$. The spectra form the data basis and are translated into changes in transmission and central frequency, both relative to the initial probe spectrum without pumping. The experimental data plotted in \autoref{fig:setup}b shows typical behaviour for a probe frequency slightly below the ENZ point: a strong initial increase of the transmission as a function of time for all three angles ($5^\circ$,$50^\circ$,$75^\circ$), up to $200$ to $300$\,\% of the initial transmission. Simultaneously, very different frequency shifting effects are measured: For the low angle case ($5^\circ$) the typical red shifting behaviour is observed, while the high angle case ($75^\circ$) presents a blue shift - as described above, this is a signature dominated by the change in phase at the interface. This interface effect can act as an opposing shift and may be utilised to tailor the frequency shift or even suppressing it entirely, as seen for $50^\circ$.

\subsection*{Comparison between experiment and model}
To compare experiment and model, we calculate the expected frequency shift by extracting the time dependent bulk plasma frequency from the measured transmission changes. We know the initial transmission and use our thin layer model to link these changes to the phase by assuming a reduction of the bulk plasma frequency. The extracted average plasma frequencies across the layer corresponding to the results shown in \autoref{fig:shift} are plotted in supplementary S3. The temporal derivative of the phase reveals the expected frequency shift seen in \autoref{fig:shift}a. Indeed, a red shift and blue shift feature are extracted and they agree quantitatively well with our measurement (see \autoref{fig:shift}b), especially considering there are no fitting parameters.

\begin{figure}[htb]
	\centering
	\includegraphics[width=0.8\columnwidth]{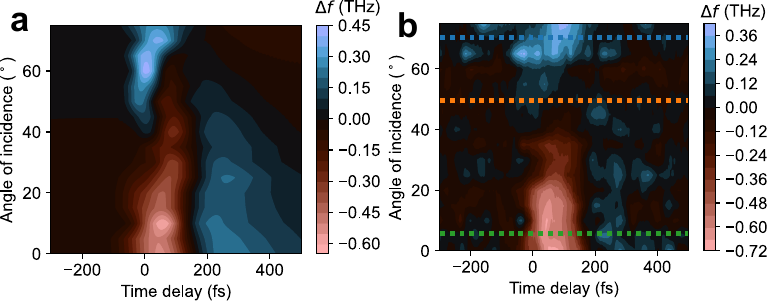}
	\caption{\label{fig:shift} \textbf{Modelling of experimental data.} For the full angle range the transmission changes are analysed and used to test the validity of our model ($I_0=400$\,GW\,cm$^{-2}$,$f_{\rm pm}=250$\,THz, $f_{\rm pr}=200$\,THz, TM analyser).  \textbf{a}, The plotted frequency shift estimates are based on the time dependent transmission phase, calculated via the previously extracted time dependent plasmon frequency. \textbf{b}, The experimentally measured frequency shifts. The measurements of \autoref{fig:setup}b are marked with dashed lines of the corresponding colours.}
\end{figure}  

\subsection*{Thickness dependence}
\begin{figure}[htb]
	\centering
	\includegraphics[width=0.8\columnwidth]{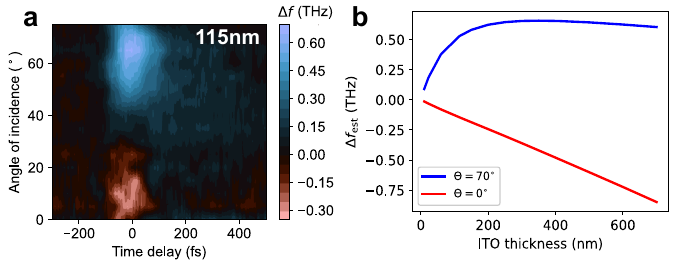}
	\caption{\label{fig:t} \textbf{Thickness dependence. a,}  The experimentally measured frequency shift of a 115\,nm layer  ($I_0=400$\,GW\,cm$^{-2}$, $f_{\rm pm}=214$\,THz, $f_{\rm pr}=240$\,THz, TM analyser). \textbf{b} , To further investigate the thickness dependence we model the ITO for the two angle extremes depending on the layer size. The quickly saturating blueshift (high angle) and a linearly increasing red-shift (low angle) are plotted. ($\omega_{\rm p}$ reduction by 10\,\%, 417\,nm sample parameters, $f_{\rm pr}=200$\,THz)}
\end{figure}

 Intuitively, one would expect the effect of interfaces to be more important for thinner samples. We demonstrate this effect experimentally by reducing the ITO layer thickness to 115\,nm thickness. \autoref{fig:t}a clearly shows a strongly reduced red shift feature, agreeing with the expectation of a reduced bulk effect. Additionally, with the decline of the red shift the blue shift feature has become more prominent. In experiment,  we observe that the frequency shift in this region is even larger than that measured for the 407\,nm thick film, an effect which is explained primarily by a lower scattering rate for this particular sample (see supplementary section S1).

 We repeat the initial simulations shown in \autoref{fig:model} for a varying thickness.  Plotted in \autoref{fig:t}b the red shift feature scales linearly with the thickness of the sample, which is inline  with expectations that the signal is predominantly determined by bulk phase change effects. However, the high angle blue shift feature saturates for film thicknesses beyond 200\,nm, where internal reflections are negligible (see Figure S3b). For films thinner up to $\sim700\,$nm, the interface dominated blue shift can be even larger than the bulk phase change effect, assuming equivalent changes to the refractive index. Hence, spatiotemporal refraction and the corresponding blue shift feature present a promising tool for controlling nonlinear frequency shifting in \textasciitilde 100\,nm to 1\,\textmu m sized samples, typical for ENZ nonlinear optics.

\section*{Conclusions}
In conclusion, we show that spatiotemporal refraction provides a frequency shift that is relatively strong compared to the temporal refraction for sub \textmu m samples.  This contribution remains constant down to film thicknesses of $\sim 100$\,nm. Most importantly we demonstrate that the frequency shift can now be controlled not only by the temporal refractive index changes, but also by the angle. This enables tailoring the nonlinear frequency shift in thin layers independent of other nonlinear modulations such as the transmission. In future, more versatile layer choices or combining multiple layers could provide a plethora of ultrafast amplitude and frequency switching devices, by tuning interface effects to obtain the desired results.

\section*{Methods}

\subsection*{Sample fabrication}
To experimentally investigate this effect we utilise ITO samples of difference thicknesses. The ITO was sputtered onto cover glass at room temperature using 90/10 In$_2$O$_3$/SnO$_2$ Kurt Lesker target and sputtering tool.  The base pressure before deposition was in low $10^{-6}$\,torr, but raised to 3\,mT of Ar only during deposition with a RF power of 145\,W.  In order to achieve high carrier density, both deposition and annealing was performed under in a lowest possible residual oxygen environment.  The samples are post-annealed in forming gas for 3\,min at temperatures between 425-525$^{\circ}$C in a rapid thermal annealer. The 407\,nm sample was obtained from UQG Ltd.

\subsection*{Optical set-up}
For the pump-probe measurements, we used an amplified Ti:sapphire laser (Legend Elite, Coherent), with a central wavelength of 800~nm, pulse duration of 107~fs and repetition rate of 1~kHz, feeding two identical OPAs (TOPAS, Light Conversion). The signal output of one OPA was used as the pump, and the signal output of the other OPA was used as the probe, allowing us independent control of pump and probe frequencies. The pump was focused using a 30~cm BK7 lens, the probe with a 25~cm CaFl$_2$ lens. The pump beam diameter (FWHM) was measured to be 480~μm in air, while the probe was 250~μm. To make sure the probes intensity is significantly smaller than the pump we used several additional OD filters to decrease the probe power and tested that the nonlinear reflection was independent of adding/removing filters. We used a referenced tuneable filter wheel to enable a frequency independent pump power.  The angle of incidence of the pump is 5$^{\circ}$ smaller than that of the probe. For the spectral analysis we used an Andor Shamrock 163 spectograph with a DU490A-1.7 camera.

\begin{backmatter}

\bmsection{Acknowledgments}
We acknowledge financial support from the Engineering and Physical Sciences Research Council (EPSRC) of the United Kingdom, via the EPSRC Centre for Doctoral Training in Metamaterials (Grant No. EP/L015331/1). TSL acknowledge the support of the U.S. Department of Energy, Office of Basic Energy Sciences, Division of Materials Sciences and Engineering. Parts of this work was performed, at the Center for Integrated Nanotechnologies, an Office of Science User Facility operated for the U.S. Department of Energy (DOE) Office of Science. We thank Philip Thomas for the ellipsometry measurement of the ITO thin film.  We thank Bill Barnes, Zahirul Alam and Simon Horsley for helpful discussions.

\bmsection{Disclosures}
The authors declare no conflicts of interest.


\bmsection{Supplemental document}
See Supplement 1 for supporting content. 

\end{backmatter}



\bibliography{references}

\end{document}


\title{Supplement 1: Spatiotemporal refraction of light in an epsilon-near-zero ITO layer}
\author{Justus Bohn\authormark{1,*},Ting Shan Luk\authormark{2,3}  and Euan Hendry\authormark{1}}

\address{\authormark{1} School of Physics, University of Exeter, Exeter, UK\\
\authormark{2}Sandia National Laboratories, Albuquerque, NM, USA\\
\authormark{3}Center for Integrated Nanotechnologies, Sandia National Laboratories, Albuquerque, NM, USA}

\email{\authormark{*}jb933@exeter.ac.uk} 

\tableofcontents

\newpage	

\section{ITO parameters}
\begin{figure}[htb]
	\centering
	\includegraphics[width=0.7\columnwidth]{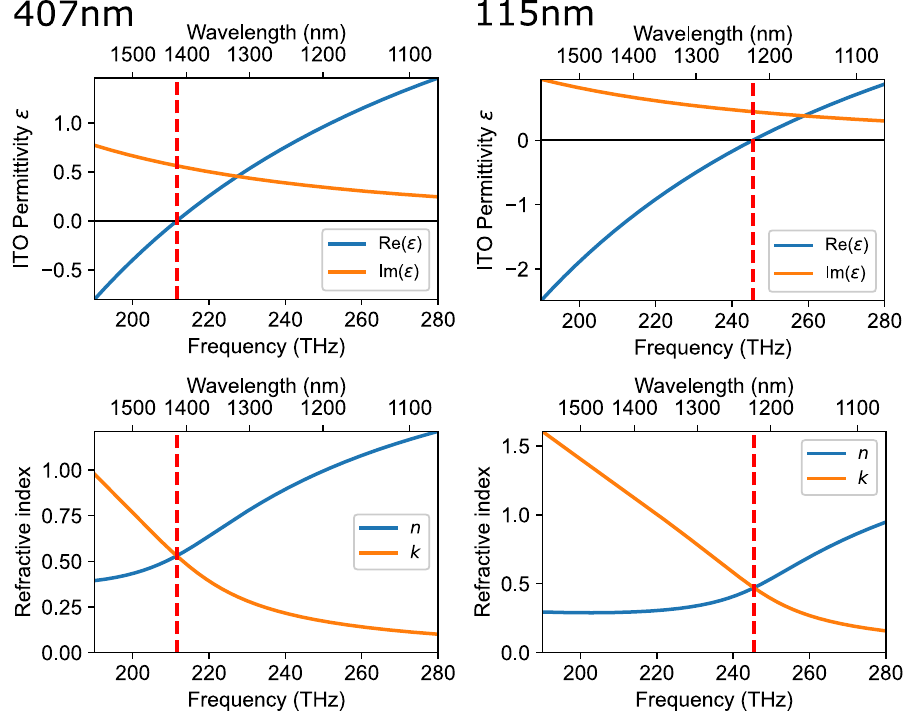}
	\caption{\label{sfig:parameters} \textbf{Optical properties of indium thin films.} Optical permittivity (top) and refractive index (bottom) are plotted for the two different samples. The ENZ frequency is marked at 212\,THz for the 407\,nm sample (left), 246\,THz for 115\,nm (right).} 
\end{figure}

The indium tin oxide (ITO) samples are characterised using an ellipsometer. The extracted optical parameters are listed in \autoref{stab:parameters}, where the epsilon-near-zero (ENZ) frequency is calculated as $f_{\rm ENZ} = \frac{1}{2\pi}\sqrt{\omega_{\rm p}^2/\varepsilon_{\infty}-\gamma^2}$. The corresponding wavelength dependent permittivity and refractive index are plotted in \autoref{sfig:parameters}.

\begin{table}[htb]
\centering
\caption{\label{stab:parameters} Drude parameters for the different ITO film thicknesses.}
\begin{tabular}{lllll}
$t_{\rm ITO}$ (nm) & $\varepsilon_{\infty}$ & $\omega_{\rm p} (10^{15}$ rad/s) & $\omega_{\rm p}/\gamma $ & $f_{\rm ENZ}$ (THz) \\ \hline
407                & 3.45                   & 2.5                              & 11.5                     & 212                 \\
115                & 3.81                   & 3.03                             & 16.9                     & 246          
\end{tabular}
\end{table}

\section{Thin layer model}
The thin layer model is based on the transfer matrix method (TMM), which simplifies to the airy formula for the three layer case (1. air, 2. ITO, 3. substrate):

\begin{equation}
\label{eq:airy}
    t = \overbrace{\frac{1}{1 + r_{12} r_{23}{\rm e}^{2{\rm i} k_2 d}}}^{\text{internal reflections}} \underbrace{t_{12} t_{23}}_{\text{interface transmission}}\overbrace{{\rm e}^{{\rm i} k_2 d}}^{\text{bulk propagation}}.
\end{equation}

The transmission coefficient ($t$) of the layer can be calculated using the sample thickness ($d=t_{\rm ITO}$), the in-plane wavevector component inside the ITO layer ($k_2 = n_{\rm ITO} k_0 \cos\Theta_2$) and the Fresnel coefficients of the front ($r_{12}$, $t_{12}$) and back interface ($r_{23}$, $t_{23}$). Here, the ${\rm e}^{{\rm i} k_2 d}$-term describes the bulk propagation term and time dependent phase changes of it have been discussed as temporal refraction. Considering the multiplication of complex numbers leads to the addition of phases it follows: 

\begin{equation}
\label{eq:ph}
    \Delta \Phi_{\rm TMM} = \overbrace{\Delta \Phi_{\rm den}+\Delta \Phi_{12}+\Delta \Phi_{23}}^{\text{interface effects}} + \overbrace{\Delta \Phi_{\rm e}}^{\text{bulk propagation}}.
\end{equation}

And the expected frequency shifts are:

\begin{equation}
\label{eq:df}
    \Delta f_{\rm TMM} = \overbrace{\Delta f_{\rm den}+\Delta f_{12}+\Delta f_{23}}^{\text{spatio-temporal refraction}} + \overbrace{\Delta f_{\rm e}}^{\text{temporal refraction}}.
\end{equation}



In this paper, we investigate the case where the temporal refraction is no longer large compared to the other frequency shift contributions. To simulate the temporal changes to the transmission amplitude and phase we introduce a time varying bulk plasmon frequency of up to 10\,\% reduction. The temporal shape is given by a convolution of a Gaussian with a full-width-half-maximum pulse length of 107\,fs and a 150\,fs exponential decay (see \autoref{sfig:m-components}a) and corresponds to the example plotted in \textbf{Figure 2} of the manuscript. Additionally, we plot the time dependent permittivity, highlighting the ENZ crossing. The temporal changes of the phase plotted in \textbf{Figure 2}b,c result in frequency shifts that are shown in \autoref{sfig:m-components}b. 

\begin{figure}[htb]
	\centering
	\includegraphics[width=0.7\columnwidth]{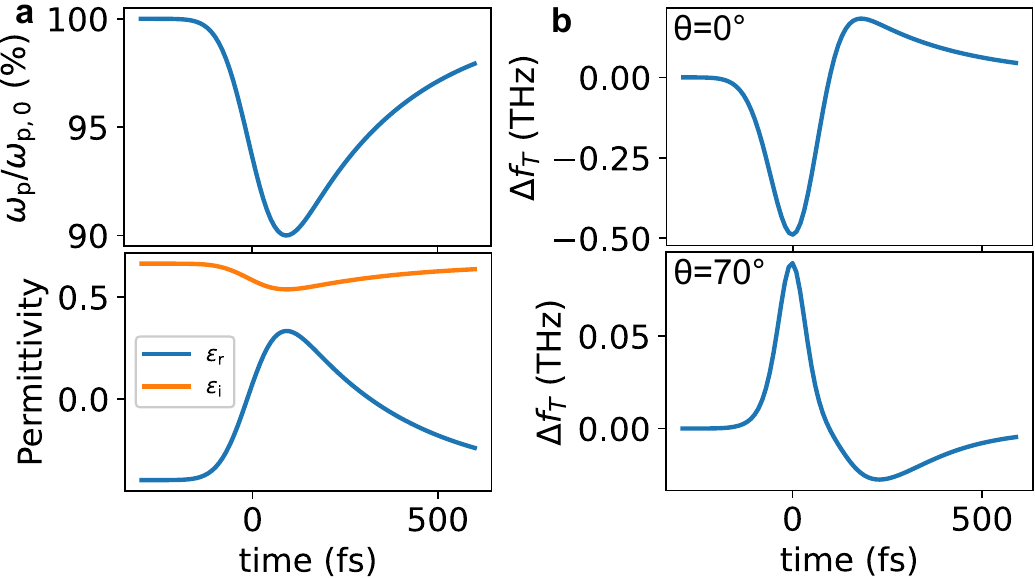}
	\caption{\label{sfig:model} \textbf{Spatiotemporal refraction: frequency shifts (corresponding to Figure 2).} We study the impact of a typical $\omega_p$ reduction by 10\,\%, as plotted \textbf{a}. The resulting permittivity is plotted for the case of the 407\,nm ITO sample parameters at 200\,THz. The ENZ case is passed at roughly maximum slope. \textbf{b}, The expected frequency shifts for the transmitted probe are calculated using $f_{\rm TMM}$ and plotted for the normal and high angle incidence case.}
\end{figure} 

The origin of these phase shifts can best be dissected by investigating the constituents of the transmittance coefficient \autoref{eq:airy}. We compare in \autoref{sfig:m-components} the total calculated transmission phase $\Phi_{\rm TMM}$ with the Fresnel transmittance coefficient phase of both interfaces ($\Phi_{12}, \Phi_{23}$) and the bulk term ($\Phi_{\rm e}=\angle({\rm e}^{{\rm i} k_2 d})$).  The normal incidence case the two interface effects cancel each other out.  However, for the high angle case the bulk contribution disappears and the front interface flips sign, while the back interface remains almost unaffected. An overall transmittance phase is now decreasing instead of increasing.

\begin{figure}[htb]
	\centering
	\includegraphics[width=0.8\columnwidth]{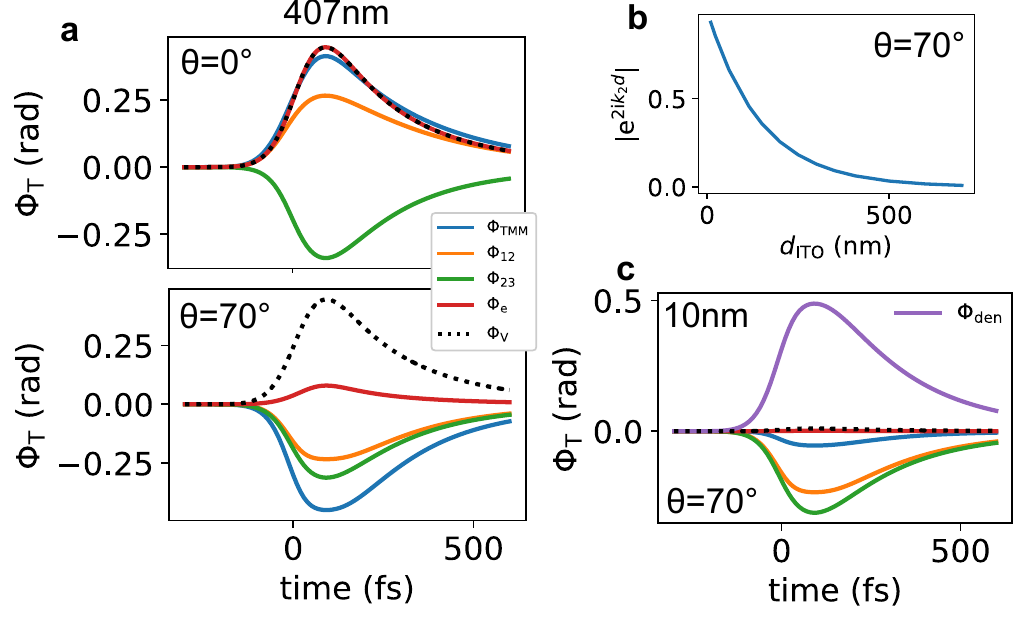}
	\caption{\label{sfig:m-components} \textbf{Spatiotemporal refraction: phase components (corresponding to Figure 2). a} The components are plotted for the low and high angle case: The total calculated transmission phase $\Phi_{\rm TMM}$, the Fresnel transmittance coefficient phase of both interfaces ($\Phi_{12}, \Phi_{23}$) and the bulk term ($\Phi_{\rm e}=\angle({\rm e}^{{\rm i} k_2 d})$). As an additional reference we plot the expected bulk contribution ($\Phi_V$) for normal incidence. \textbf{b}, We plot the absolute value of the exponential term in the denominator, showing that it only becomes comparable to 1, and therefore relevant, for $d\lesssim200$\,nm. \textbf{c}, For the very thin case of 10\,nm (${\rm e}^{2{\rm i} k_2 d}\sim1$) the reflection based phase shifts caused by the denominator becomes large and compensates transmission based effects.}
\end{figure} 

Going to small thicknesses ($d\ll1/2\,{\rm Im}(k_2)$) the multiple reflection based denominator becomes relevant (see \autoref{sfig:m-components}b)  and the phase shift of the internal reflection compensate the transmission based effects (see \autoref{sfig:m-components}c). Hence, the blue shift is expected to decreases for thicknesses smaller than $\sim200\,$nm as seen in \textbf{Figure 5}b of the manuscript.

\subsection{Frequency dependence}

\begin{figure}[htb]
	\centering
	\includegraphics[width=0.8\columnwidth]{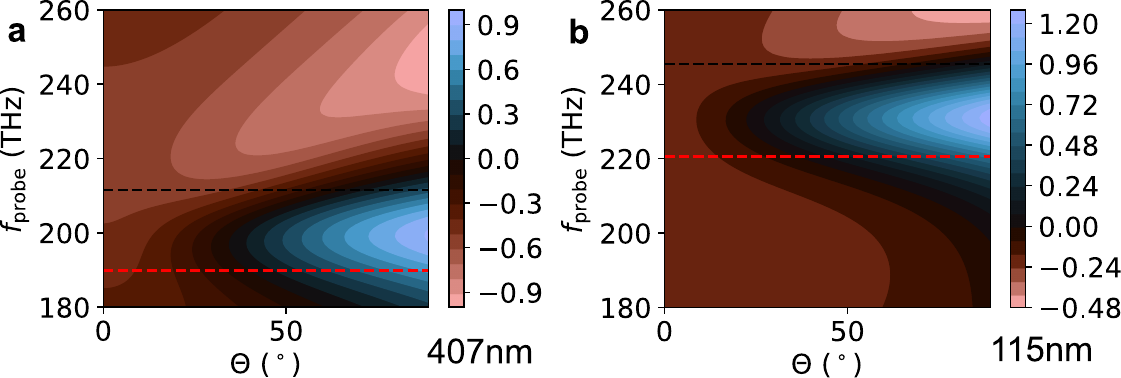}
	\caption{\label{sfig:m-global} \textbf{Angle and frequency dependent spatio-temporal refraction behaviour.} We plot the frequency shifts as seen in \autoref{sfig:model}b for the maximum case of $\tau=0\,$fs and study the probe frequency and angle dependence for the case of the 407\,nm (\textbf{a}) and the 115\,nm (\textbf{b}) parameters. The black dashed line indicates the initial $f_{\rm ENZ}$, while the red dashed line indicates the pumped case.}
\end{figure}

To study the frequency shifting behaviour over large frequency and angle ranges, we investigate the maximum frequency shift exclusively (see \autoref{sfig:m-global}). For both, the 407\,nm and 115\,nm sample, the blue shifted feature appears only for large angle of incidence, as established earlier. Additionally, this blue shift occurs only for frequencies just below the (initial) ENZ case. In fact, the region of maximum blue shift sits right between the initial and pumped ENZ frequency case. This can be understood by considering the Fresnel coefficient of the front interface, which dominantly affects this feature as seen in \autoref{sfig:m-components}a. For the ENZ case  Taylor expansion slightly below $90^\circ$ the coefficient is given as ($\Theta\lessapprox\pi/2$, $\Theta_2\ll\Theta\xrightarrow{}n_{\rm ITO} \ll 1$, TM-polarization): 

\begin{equation} 
    t_{\rm 12}(\tau) \propto \sqrt{\frac{\varepsilon_{\rm ITO}(\tau)}{\varepsilon_{\rm ITO}(\tau)-1}}\stackrel{|\varepsilon_{\rm ITO}| \ll 1}{\approx} {\rm i}\sqrt{\varepsilon_{\rm ITO}(\tau)}.
\end{equation}


Hence, $t_{\rm 12}$  reduces in phase upon crossing the ENZ case (starting from negative $\varepsilon_{\rm r}$ values as seen in \autoref{sfig:model}a).

\newpage

\section{Applying the model to experimental data}

\begin{figure}[ht!]
	\centering
	\includegraphics[width=0.8\columnwidth]{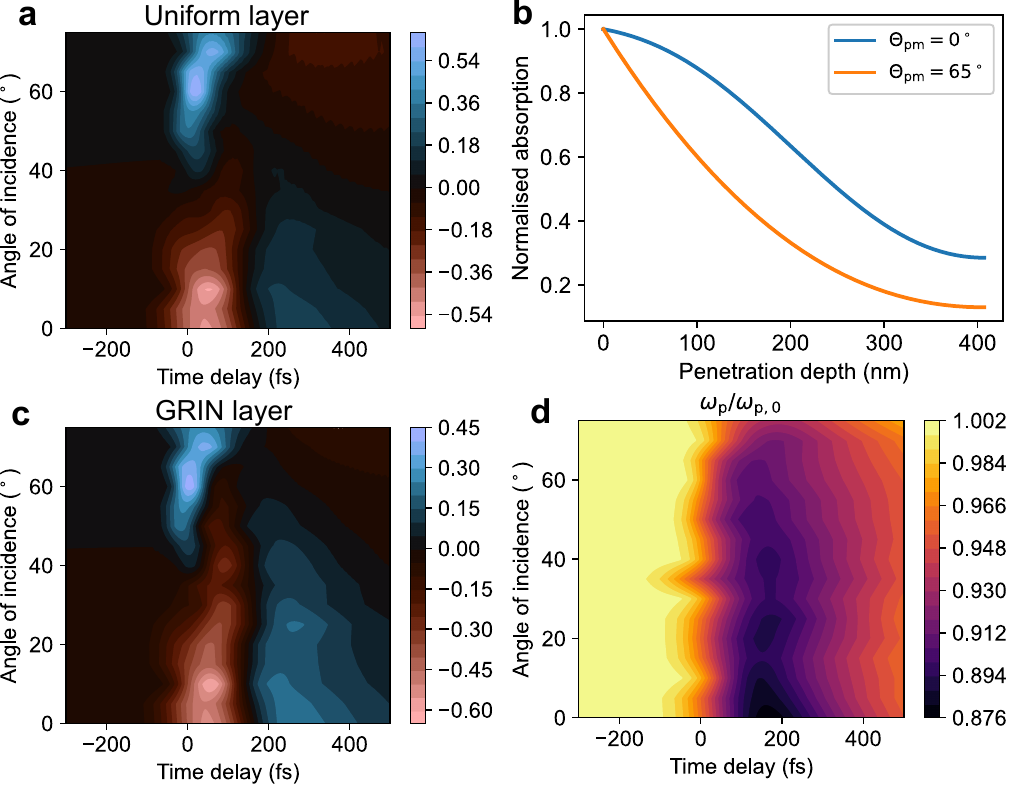}
	\caption{\label{sfig:m-exp} \textbf{Uniform vs GRIN layer (corresponding to Figure 4). a,} The plotted frequency shift estimates are based on the time dependent transmission phase, calculated via the uniform layer assumption ($I_0=400$\,GW\,cm$^{-2}$,$f_{\rm pm}=250$\,THz, $f_{\rm pr}=200$\,THz, TM analyser).  \textbf{b}, The normalised depth dependent absorption  of the TE polarised pump for the low and high angle case plotted in Figure 3b. \textbf{c}, The GRIN based frequency shift estimates, utilising the depth dependent plasma frequency. \textbf{d}, The plasma frequencies plotted correspond to the resulting transmission changes measured in the experiment corresponding to Figure 4b.}
\end{figure}

To test the model we put the implicit assumption under test that the transmission changes and phase changes have the same origin, a red-shift in $\omega_{\rm p}$. The transmission changes are linked to the red-shifting of $\omega_{\rm p}$. These can be used to calculate all-optical parameter changes of interest. In \autoref{sfig:m-exp}a we show the case of using a uniform ITO layer excitation and find that especially the blue-shift seems to be overestimated. While the temporal dynamics are much more complicated then this, we attempt to get a slightly better estimate by assuming a spatial dependence of the $\omega_{\rm p}$ in ITO. This results in a graded refractive index (GRIN) profile. We use the transfer matrix method to calculate the depth dependent absorption of the pump inside the ITO layer as seen in \autoref{sfig:m-exp}b and scale the distribution. The front will be most affected by pumping and $t_{12}$ driven effects may saturate sooner, while the overall blue-shift may be smaller as the angle independent $t_{23}$  contribution (see \autoref{sfig:m-components}a) becomes smaller. This can be seen in \autoref{sfig:m-exp}c of our model application, which behaves still qualitatively very similar to the uniform layer plotted in a, but has a significantly smaller blue shift. This also corresponds better to what is measured in experiment. While this depth dependent plasma frequency should still be seen as a rough assumption, it appears to be noticeably more accurate than simply taking a uniform layer. The extracted $\omega_{\rm p}$ values averaged over the layer thickness are plotted in \autoref{sfig:m-exp}d. The modulation of up to $\sim10\,$\% are in good agreement with similar studies \cite{Bohn2021AllopticalSwitchingEpsilonnearzero}. Also, the increased angle leads to a reduced intensity due to a streched out elliptical beam-sample cross-section area, hence, a smaller bulk plasma modulation is seen for higher angles.

\newpage

\section{Additional measurements}
 Here, we present further experimental data to support the previously discussed scenarios.

\subsection{Wavelength dependence}

\begin{figure}[ht!]
	\centering
	\includegraphics[width=0.6\columnwidth]{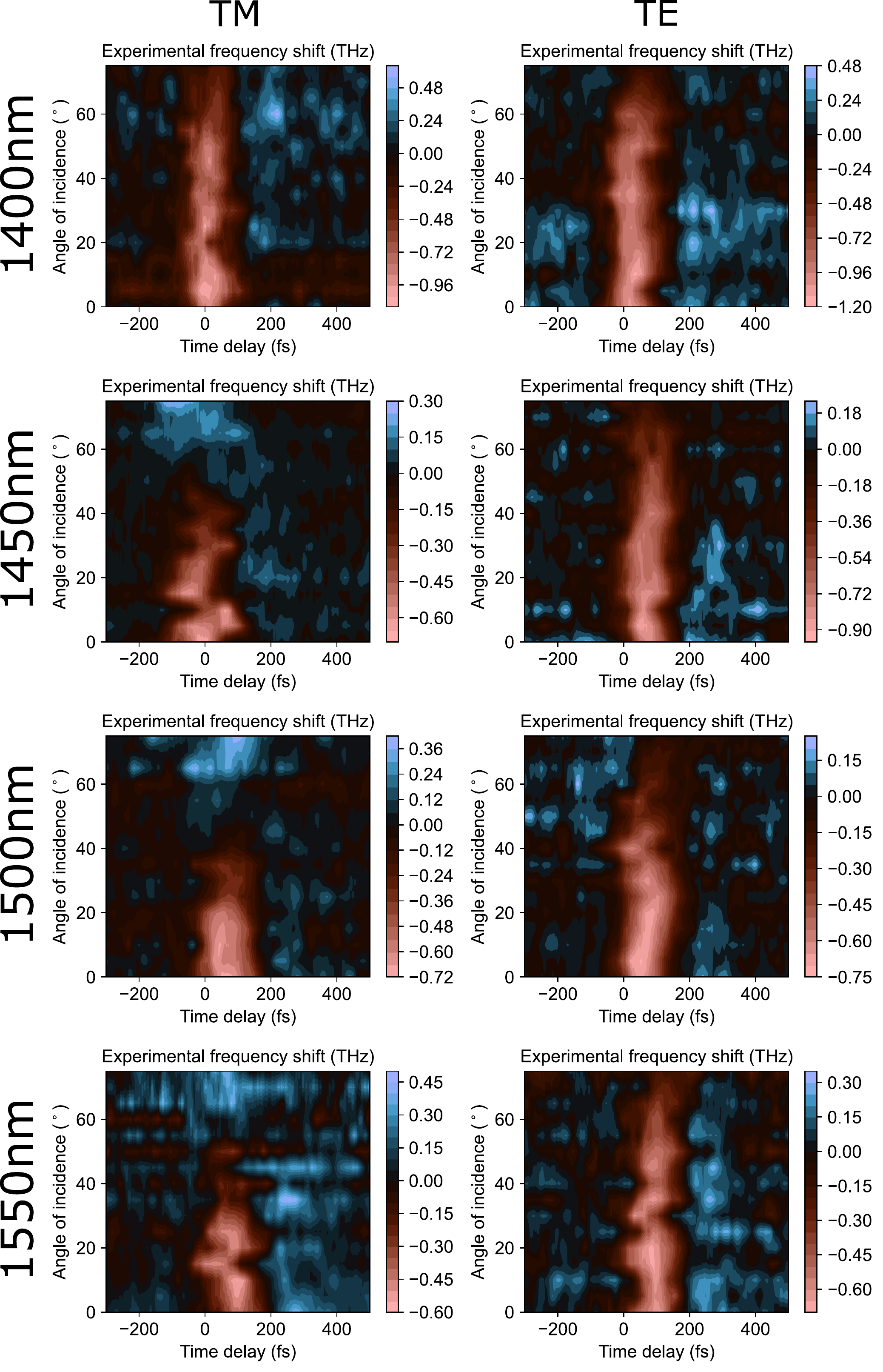}
	\caption{\label{sfig:wav-pol} \textbf{Wavelength and polarisation dependence.} The  measured frequency shifts show the wavelength dependence. Additionally, the TE case is measured by rotating the analyser. ($I_0=400$\,GW\,cm$^{-2}$,$f_{\rm pm}=250$\,THz).}
\end{figure}

The ENZ wavelength of the 407\,nm sample is $\lambda_{\rm ENZ}=1417$\,nm. We show in \autoref{sfig:wav-pol} that the blue shift does not appear above the ENZ case. Additionally we show that it is reserved for the TM polarisation.

\newpage
\subsection{Intensity dependence}
\begin{figure}[ht!]
	\centering
	\includegraphics[width=0.6\columnwidth]{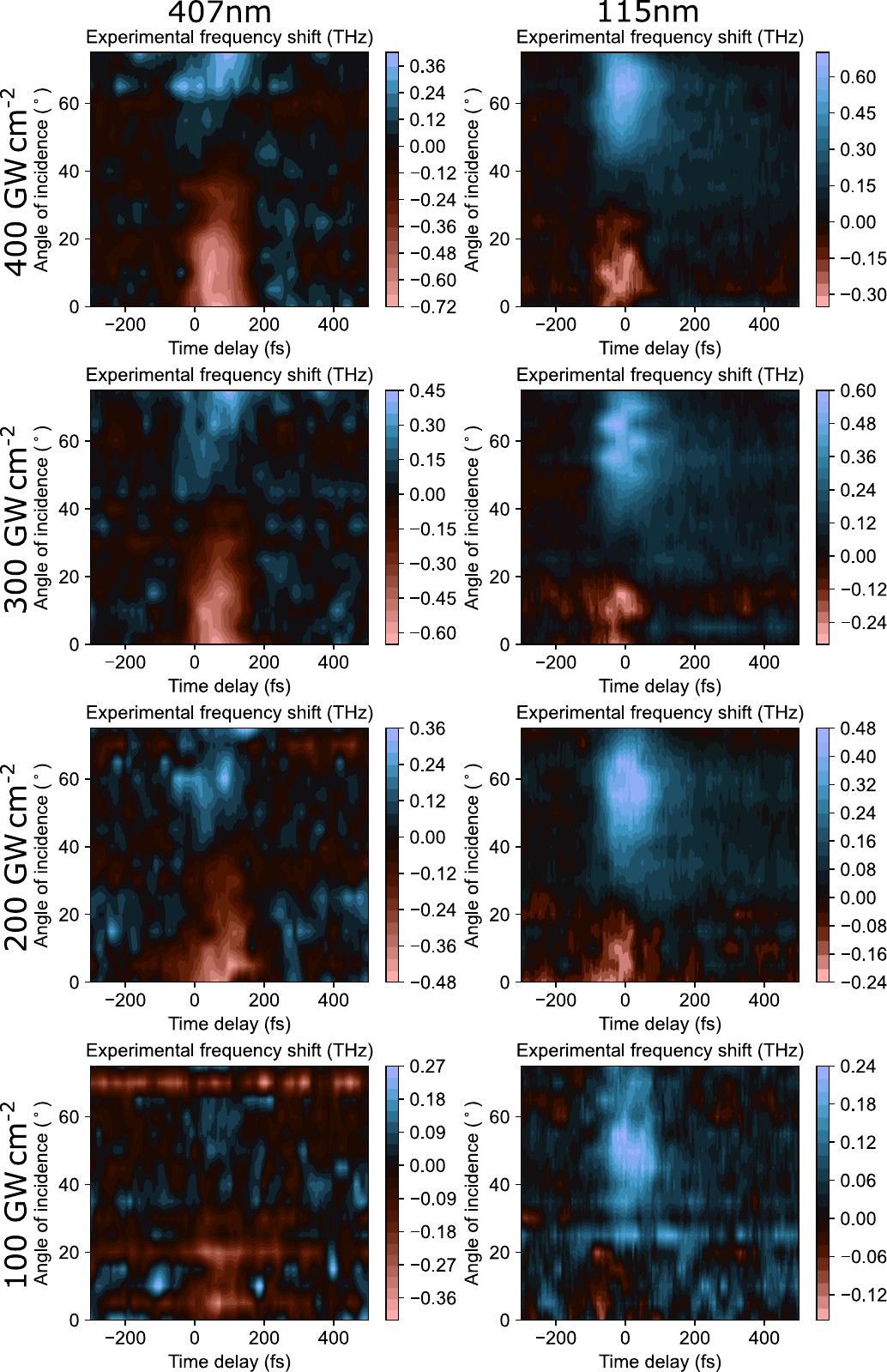}
	\caption{\label{sfig:t-int} \textbf{Intensity dependence.} The measured frequency shifts show the intensity dependence of the TM contribution. This was done for both samples: 407\,nm ($f_{\rm pm}=250$\,THz, $f_{\rm pr}=200$\,THz) and 115\,nm ($f_{\rm pm}=214$\,THz, $f_{\rm pr}=240$\,THz).}
\end{figure}

The intensity dependent measurements are done for 100 to 400\,GW\,cm$^-2$ for both samples and plotted in \autoref{sfig:t-int}. The red-shift feature appears to saturate slightly faster for the 115\,nm sample then for the 407\,nm. The blue shift feature appears to start saturating for both samples roughly at 200\,GW\,cm$^-2$.

\newpage


\subsection{Blue shifted spectrum}
\begin{figure}[htb]
	\centering
	\includegraphics[width=0.7\columnwidth]{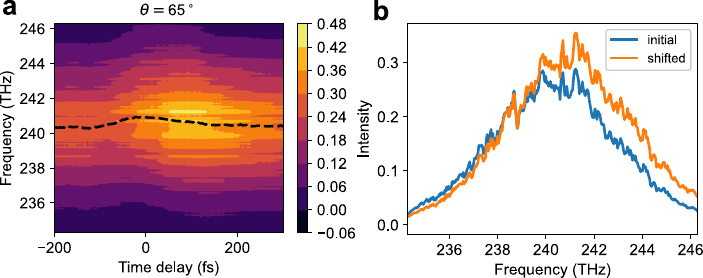}
	\caption{\label{sfig:spec} \textbf{Blue shifted spectrum (corresponding to \textbf{Figure 5}a). a}, For the 115\,nm sample the time dependent spectrum was measured. The pump intensity is 400\,GW\,cm$^-2$. The dashed line indicates the traced central frequency position.  \textbf{b}, The pump free initial spectrum (blue) is plotted along the maximum shifted case (orange).}
\end{figure}

We present the example of the strongest measured blue shift feature, given by the 115\,nm film as seen in \autoref{sfig:spec} and shown in \textbf{Figure 5}a of the manuscript.

\bibliography{references}